\documentclass{elsart}
\usepackage[dvips]{graphicx}

\usepackage{color}
\usepackage{amssymb}
\usepackage{amsmath}
\usepackage{epsf}
\usepackage{subfigure}
\begin{document}
\begin{frontmatter}
  \title{Ensemble inequivalence in random graphs}
  \author{Julien Barr\'e$^{1}$}, \author{Bruno Gon\c{c}alves $^{2}$}
   \address{1. Laboratoire J.-A.~Dieudonn\'e, Universit\'e de Nice-Sophia 
    Antipolis\\  Parc Valrose, 06108 Nice Cedex 02, France}
  \address{2. Physics Department, Emory University\\Atlanta, Georgia 30322,USA}
\thanks[julien]{Corresponding author. Tel: (+33) 4 92 07 62 34; Fax: (+33) 4 
93 51 79 74; E-mail: jbarre@unice.fr}
\thanks[bruno]{E-mail: bgoncalves@physics.emory.edu}

\begin{abstract}
  We present a complete analytical solution of a system of Potts spins
  on a random $k$-regular graph in both the canonical and
  microcanonical ensembles, using the Large Deviation Cavity Method
  (LDCM). The solution is shown to be composed of three different
  branches, resulting in an non-concave entropy function.The
  analytical solution is confirmed with numerical Metropolis and
  Creutz simulations and our results clearly demonstrate the presence
  of a region with negative specific heat and, consequently, ensemble
  inequivalence between the canonical and microcanonical ensembles.
\end{abstract}
\end{frontmatter}

\section{Introduction}
\label{sec:intro}

When a system phase-separates, it pays for the different domains with a
surface energy, which is usually negligible with respect to the bulk 
energy. As a consequence, any non concave region in 
the entropy vs energy curve has to be replaced by a straight line.
This is the result of the usual Maxwell construction. 

However, the condition of negligible surface energy is violated in
presence of long range interactions, as well as for systems with a
small number of components. In both cases, the possibility of non
concave entropies and ensemble inequivalence is well known, and has
been demonstrated on numerous models, for
instance~\cite{thirring,gross,cohen,beg}. The same condition of
negligible surface energy is also violated on sparse random graphs:
despite the fact that each site has only a small number of neighbors,
there will be in general an extensive number of links between two
(extensive) subsets of the system. The possibility of ensemble
inequivalence in this type of models has been alluded to in some works
related to the statistical physics of random graphs and combinatorial
optimization~\cite{monasson}. However, these
  authors study the analog of the canonical ensemble, and replace
  the non concave part of the entropy by a straight line. This
phenomenon remains thus to our knowledge unstudied, despite the
widespread current interest in complex interaction structures, and
networks in general. The purpose of this work is to present a simple,
exactly solvable model on a random regular network, that displays a
non concave entropy and ensemble inequivalence. This is a first step
towards the study of more complicated networks, which may also include
some local structure, like small world networks.

The paper is organized as follows: in section~\ref{sec:analytical}, we
present the model, and give its analytical solution; we then turn in
section~\ref{sec:numerical} to the comparison with microcanonical
simulations using both Creutz~\cite{creutz} microcanonical dynamics and 
Metropolis~\cite{metropolis} canonical simulations. The final section is 
devoted to conclusions and perspectives.

\section{Presentation of the model and analytical solution}
\label{sec:analytical}

\subsection{The model}

We study a ferromagnetic system of Potts spins\index{Potts spins} with three
possible states (\textbf{$a$}, \textbf{$b$} and \textbf{$c$}). 
The Hamiltonian is chosen to be:

\[
\mathcal{H}=J\sum_{\langle i,j\rangle}\left(1- \delta_{q_{i}q_{j}}\right)\]

where $\langle i,j\rangle$ denotes all the bonds in the system,
$q_{i}$ is the state of spin $i$, and $\delta_{q_{i}q_{j}}$ is a
Kronecker delta. In this form, the Hamiltonian simply counts the number of
bonds between spins in different states. The ground state energy
is~$0$. The spins are located on the nodes of a regular 
random graph where each node has connectivity~$k$, of
order~$1$. A mean field like version of this
model, with an all-to-all coupling, has been studied by Ispolatov and
Cohen~\cite{cohen}, and displays ensemble inequivalence.

\subsection{Analytical solution}
Random regular graphs possess very few loops of size order~$1$, and
locally look like trees; this feature allows us to use standard
statistical physics methods, originally developed for Bethe lattices.
These calculations are usually done in the canonical
  ensemble only; in contrast, we are interested also in the
  microcanonical solution.  We compute here the free energy and the entropy
of the system, by following the formalism of the Large Deviation
Cavity Method described by O. Rivoire in~\cite{rivoire}.  We consider
however only large deviation functions with respect to spin disorder,
and not with respect to disorder in the graph structure like
in~\cite{rivoire}.

\begin{figure}
\hskip22pt\includegraphics[clip,height=4.5cm]{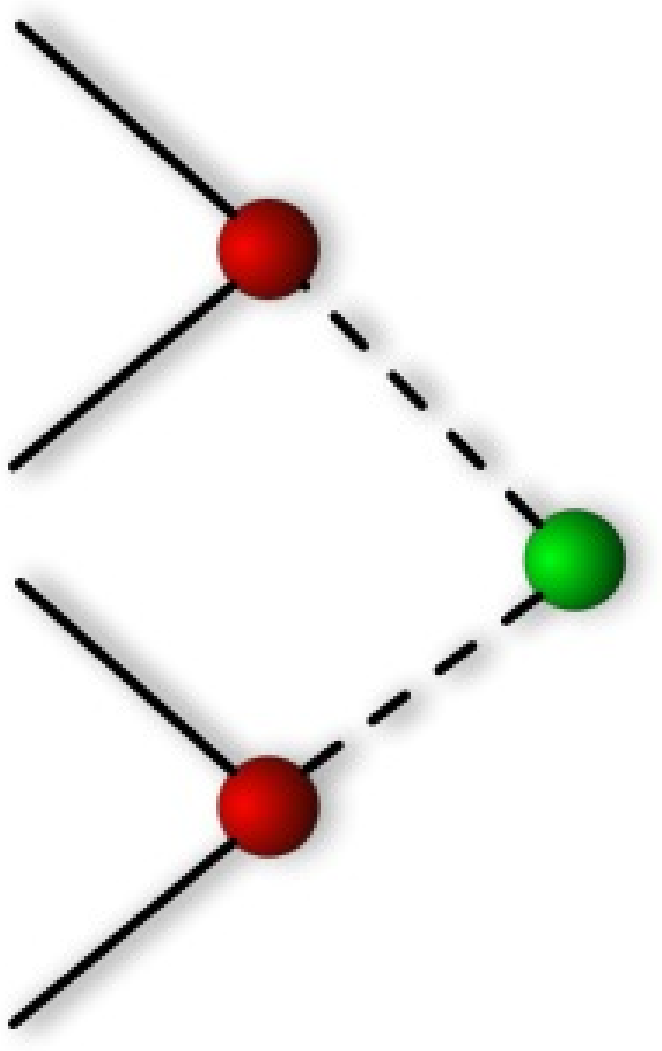}
\hskip44pt\includegraphics[clip,height=4.5cm]{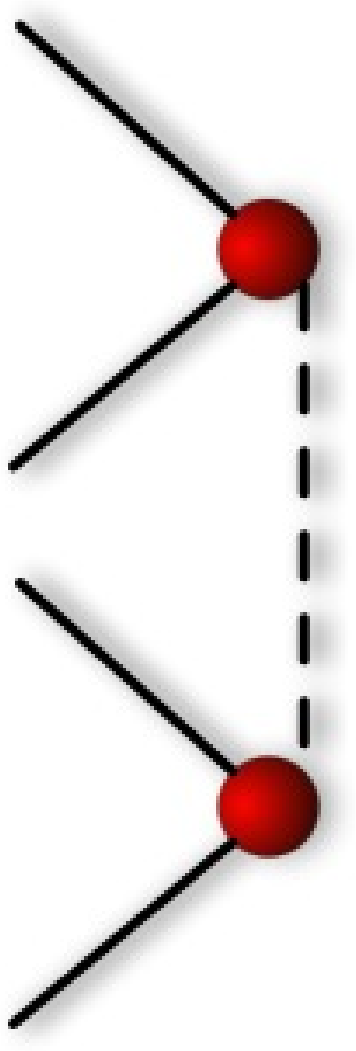}\hskip44pt
\includegraphics[clip,height=4.5cm]{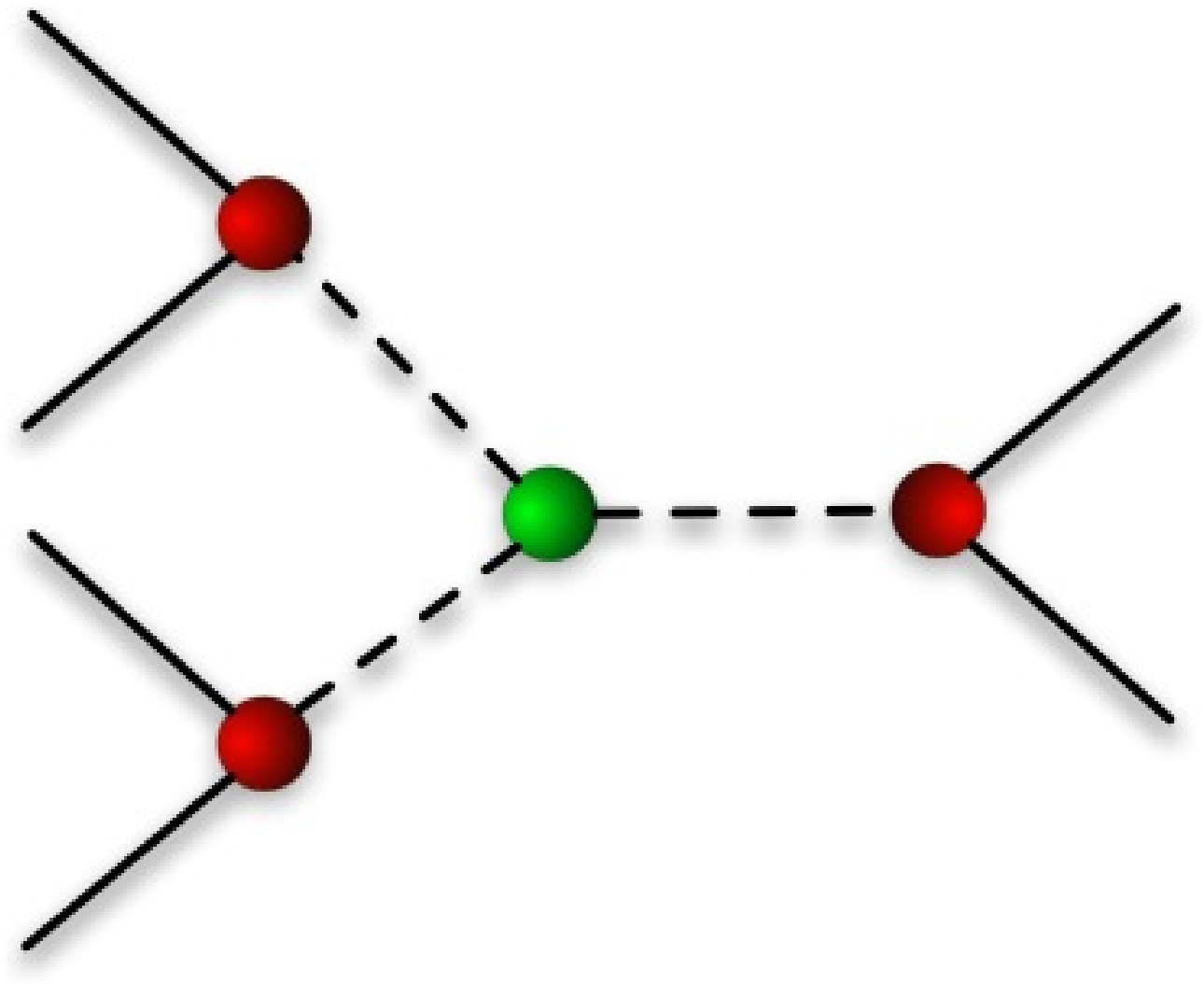}
\caption{\label{fig:iteration}Schematic representation of the
  iteration (left), link addition (center) and site addition
  (right).Red nodes and solid edges represent the original cavity
  spins and links, while the green colored nodes and dashed lines
  identify the additions.}
\end{figure}

We call cavity sites sites which have only $k-1$ neighbors, and one 
free link. Cavity site $i$ sends a field $h_i$ along each link, 
which tells its state $a$, $b$ or $c$. These field are distributed according 
to the probability distribution $P\left(h\right)$:
\begin{equation}
P\left(h\right)=p_a\delta_{h,a}+p_b\delta_{h,b}+p_c\delta_{h,c}~.
\end{equation}

\begin{table}[th]
\begin{centering}
\begin{tabular}{|c|c|c|c|}
\hline 
$h_{0}$&
$\left(h_{1},h_{2}\right)$&
$\Delta E_{n}$&
$prob$\tabularnewline
\hline
&
$\left(a,a\right)$&
$0$&
$\frac{1}{3}p_{a}^{2}$\tabularnewline
&
$\left(a,b\right)$&
$1$&
$\frac{1}{3}2p_{a}p_{b}$\tabularnewline
$a$&
$\left(a,c\right)$&
$1$&
$\frac{1}{3}2p_{a}p_{c}$\tabularnewline
&
$\left(b,b\right)$&
$2$&
$\frac{1}{3}p_{b}^{2}$\tabularnewline
&
$\left(b,c\right)$&
$2$&
$\frac{1}{3}2p_{b}p_{c}$\tabularnewline
&
$\left(c,c\right)$&
$2$&
$\frac{1}{3}p_{c}^{2}$\tabularnewline
\hline 
&
$\left(b,b\right)$&
$0$&
$\frac{1}{3}p_{b}^{2}$\tabularnewline
&
$\left(b,a\right)$&
$1$&
$\frac{1}{3}2p_{b}p_{a}$\tabularnewline
$b$&
$\left(b,c\right)$&
$1$&
$\frac{1}{3}2p_{b}p_{c}$\tabularnewline
&
$\left(a,a\right)$&
$2$&
$\frac{1}{3}p_{a}^{2}$\tabularnewline
&
$\left(a,c\right)$&
$2$&
$\frac{1}{3}2p_{a}p_{c}$\tabularnewline
&
$\left(c,c\right)$&
$2$&
$\frac{1}{3}p_{c}^{2}$\tabularnewline
\hline
&
$\left(c,c\right)$&
$0$&
$\frac{1}{3}p_{c}^{2}$\tabularnewline
&
$\left(c,a\right)$&
$1$&
$\frac{1}{3}2p_{c}p_{a}$\tabularnewline
$c$&
$\left(c,b\right)$&
$1$&
$\frac{1}{3}2p_{c}p_{b}$\tabularnewline
&
$\left(a,a\right)$&
$2$&
$\frac{1}{3}p_{a}^{2}$\tabularnewline
&
$\left(a,b\right)$&
$2$&
$\frac{1}{3}2p_{a}p_{b}$\tabularnewline
&
$\left(b,b\right)$&
$2$&
$\frac{1}{3}p_{b}^{2}$\tabularnewline
\hline
\end{tabular}
\par\end{centering}
\caption{\label{cap:Potts-iter} Analysis of the iteration process for 
$k=3$: energy shifts and probabilities. $h_0$ is the field sent by the 
new cavity site.}
\end{table}

The first step is to obtain a self consistent equation for the
probabilities $p_a,~p_b$ and $p_c$ through the analysis of the
``iteration'' process, represented on the left side of
Fig~\ref{fig:iteration}.  During an iteration step, a new site is
connected to $k-1$ cavity sites to become a new cavity site. Several
possibilities must be accounted for, corresponding to all the possible
configurations along the newly created edges.  Let us note that for
infinite temperature, or $\beta=0$, each new spin has probability
$1/3$ to be in each of the three states $a$, $b$ and $c$. This is the
origin of the $1/3$ factors in table~\ref{cap:Potts-iter} where we
represent all the terms to be considered in the $k=3$ case.

Using this table and following~\cite{rivoire}, we obtain:
\begin{equation}
\left\{ \begin{array}{l}
p_{a}=\frac{1}{Z}\frac{1}{3}\left\{ p_{a}^{2}+2p_{a}\left(p_{b}+p_{c}\right)e^{-\beta}+\left(p_{b}+p_{c}\right)^{2}e^{-2\beta}\right\} \\
p_{b}=\frac{1}{Z}\frac{1}{3}\left\{ p_{b}^{2}+2p_{b}\left(p_{a}+p_{c}\right)e^{-\beta}+\left(p_{a}+p_{c}\right)^{2}e^{-2\beta}\right\} \\
p_{c}=\frac{1}{Z}\frac{1}{3}\left\{ p_{c}^{2}+2p_{c}\left(p_{a}+p_{b}\right)e^{-\beta}+\left(p_{a}+p_{b}\right)^{2}e^{-2\beta}\right\} \\
Z=\frac{1}{3}\left\{ \left[p_{a}+\left(p_{b}+p_{c}\right)e^{-\beta}\right]^{2}+\left[p_{b}+\left(p_{a}+p_{c}\right)e^{-\beta}\right]^{2}+\left[p_{c}+\left(p_{a}+p_{b}\right)e^{-\beta}\right]^{2}\right\} \end{array}\right.\label{eq:pabc}
\end{equation}

from where we can easily calculate numerically $p_{a,b,c}$. For larger
$k$ the generalization is straightforward, we have:
\begin{equation}
p_{a}=\frac{1}{3Z}\left[p_{a}+\left(p_{b}+p_{c}\right)e^{-\beta}\right]^{k-1}
\label{eq:pabck}
\end{equation}
We compute the generalized free energy
$\mathcal{F}\left(\beta\right)$ through the formula:
\begin{equation}
\mathcal{F}\left(\beta\right)= -\ln\left[ \langle e^{-\beta\Delta
    E_{site}} \rangle\right] +\frac{k}{2}\ln\left[\langle
  e^{-\beta\Delta E_{link}} \rangle \right]~.
\label{eq:F}
\end{equation}
where $\Delta E_{site}$ and $\Delta E_{link}$ are the energy shifts
due to a site and a link addition respectively. The
$\langle~.~\rangle$ symbol denotes the expected value. Link and site
additions are depicted on the center and right sides of
Fig.~\ref{fig:iteration}, respectively. The analysis of the energy
shifts in the $k=3$ case is detailed in Tables.~\ref{cap:Potts-link}
and~\ref{cap:Potts-site}.

\begin{table}[th]
\begin{centering}
\begin{tabular}{|c|c|c|c|}
\hline 
$\left(h_{1},h_{2}\right)$&
$\Delta E$&
proba.&
$P_{l}\left(\Delta E\right)$\tabularnewline
\hline
$\left(a,a\right)$&
$0$&
$p_{a}^{2}$&
\tabularnewline
$\left(b,b\right)$&
$0$&
$p_{b}^{2}$&
$p_{a}^{2}+p_{b}^{2}+p_{c}^{2}$\tabularnewline
$\left(c,c\right)$&
$0$&
$p_{c}^{2}$&
$ $\tabularnewline
\hline 
$\left(a,b\right)$&
$1$&
$2p_{a}p_{b}$&
$ $\tabularnewline
$\left(a,c\right)$&
$1$&
$2p_{a}p_{c}$&
$2\left(p_{a}p_{b}+p_{a}p_{c}+p_{b}p_{c}\right)$\tabularnewline
$\left(b,c\right)$&
$1$&
$2p_{b}p_{c}$&
$ $\tabularnewline
\hline
\end{tabular}
\par\end{centering}
\caption{\label{cap:Potts-link} Configurations $\left(h_{1},h_{2}\right)$,
energy shifts $\Delta E$ and total probabilities $P_{l}\left(\Delta E\right)$
for the case of a link addition. The numeric factors stem from combinatoric
arguments.}
\end{table}

\begin{table}[th]
\begin{centering}
\begin{tabular}{|c|c|c|c|}
\hline 
\mbox{new site}&
$\left(h_{1},h_{2},h_{3}\right)$&
$\Delta E$&
$P_{n}\left(\Delta E\right)$\tabularnewline
\hline
\hline 
&
$\left(a,a,a\right)$&
$0$&
$\frac{1}{3}p_{a}^{3}$\tabularnewline
$a$&
$\left(a,a,b\right),\left(a,a,c\right)$&
$1$&
$\frac{1}{3}\left(3p_{a}^{2}p_{b}+3p_{a}^{2}p_{c}\right)$\tabularnewline
&
$\left(a,b,b\right),\left(a,b,c\right),\left(a,c,c\right)$&
$2$&
$\frac{1}{3}\left(3p_{a}p_{b}^{2}+3p_{a}p_{c}^{2}+6p_{a}p_{b}p_{c}\right)$\tabularnewline
&
$\left(b,b,b\right),\left(b,b,c\right),\left(b,c,c\right),\left(c,c,c\right)$&
$3$&
$\frac{1}{3}\left(p_{b}^{3}+p_{c}^{3}+3p_{b}p_{c}^{2}+3p_{c}p_{b}^{2}\right)$\tabularnewline
\hline
&
$\left(b,b,b\right)$&
$0$&
$\frac{1}{3}p_{b}^{3}$\tabularnewline
$b$&
$\left(b,b,a\right),\left(b,b,c\right)$&
$1$&
$\frac{1}{3}\left(3p_{b}^{2}p_{a}+3p_{b}^{2}p_{c}\right)$\tabularnewline
&
$\left(b,a,a\right),\left(b,a,c\right),\left(b,c,c\right)$&
$2$&
$\frac{1}{3}\left(3p_{b}p_{a}^{2}+3p_{b}p_{c}^{2}+6p_{b}p_{a}p_{c}\right)$\tabularnewline
&
$\left(a,a,a\right),\left(a,a,c\right),\left(a,c,c\right),\left(c,c,c\right)$&
$3$&
$\frac{1}{3}\left(p_{a}^{3}+p_{c}^{3}+3p_{a}p_{c}^{2}+3p_{c}p_{a}^{2}\right)$\tabularnewline
\hline
\multicolumn{1}{|c|}{}&
$\left(c,c,c\right)$&
$0$&
$\frac{1}{3}p_{c}^{3}$\tabularnewline
\multicolumn{1}{|c|}{$c$}&
$\left(c,c,b\right),\left(c,c,a\right)$&
$1$&
$\frac{1}{3}\left(3p_{c}^{2}p_{b}+3p_{c}^{2}p_{a}\right)$\tabularnewline
\multicolumn{1}{|c|}{}&
$\left(c,b,b\right),\left(c,b,a\right),\left(c,a,a\right)$&
$2$&
$\frac{1}{3}\left(3p_{c}p_{b}^{2}+3p_{c}p_{a}^{2}+6p_{c}p_{b}p_{a}\right)$\tabularnewline
&
$\left(b,b,b\right),\left(b,b,a\right),\left(b,a,a\right),\left(a,a,a\right)$&
$3$&
$\frac{1}{3}\left(p_{b}^{3}+p_{a}^{3}+3p_{b}p_{a}^{2}+3p_{a}p_{b}^{2}\right)$\tabularnewline
\hline
\end{tabular}
\par\end{centering}
\caption{\label{cap:Potts-site}Possible configurations 
$\left(h_{1},h_{2},h_{3}\right)$,
energy shifts $\Delta E$ and probabilities for the different states
in which the new site can be. The overall factor of $\frac{1}{3}$
corresponds to the \emph{a priori} probability that the new site is
in state $a$ and the remaining numeric multipliers stem from combinatorics. }
\end{table}

\begin{figure}
\includegraphics[clip,width=7cm]{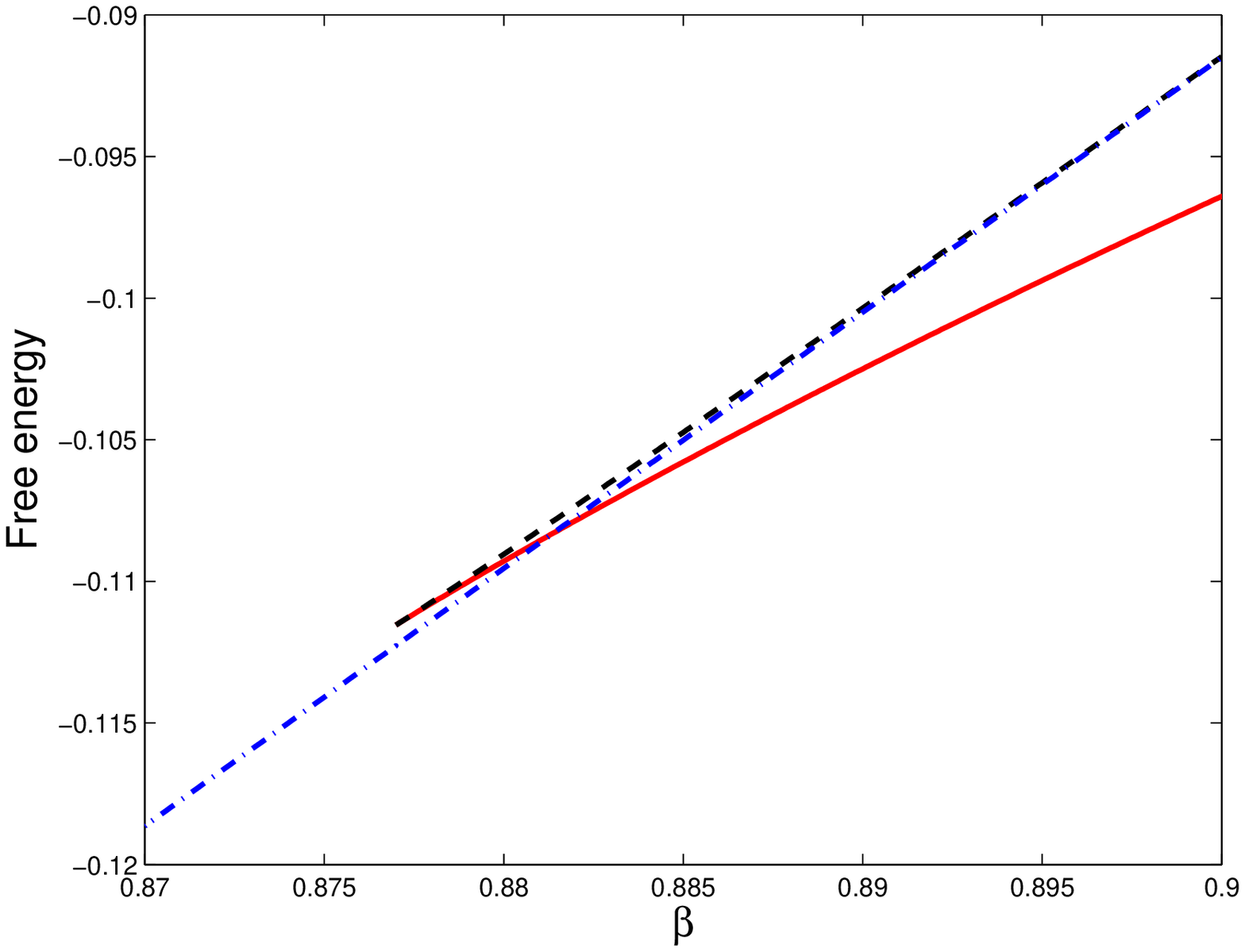}\includegraphics[clip,width=7cm]
{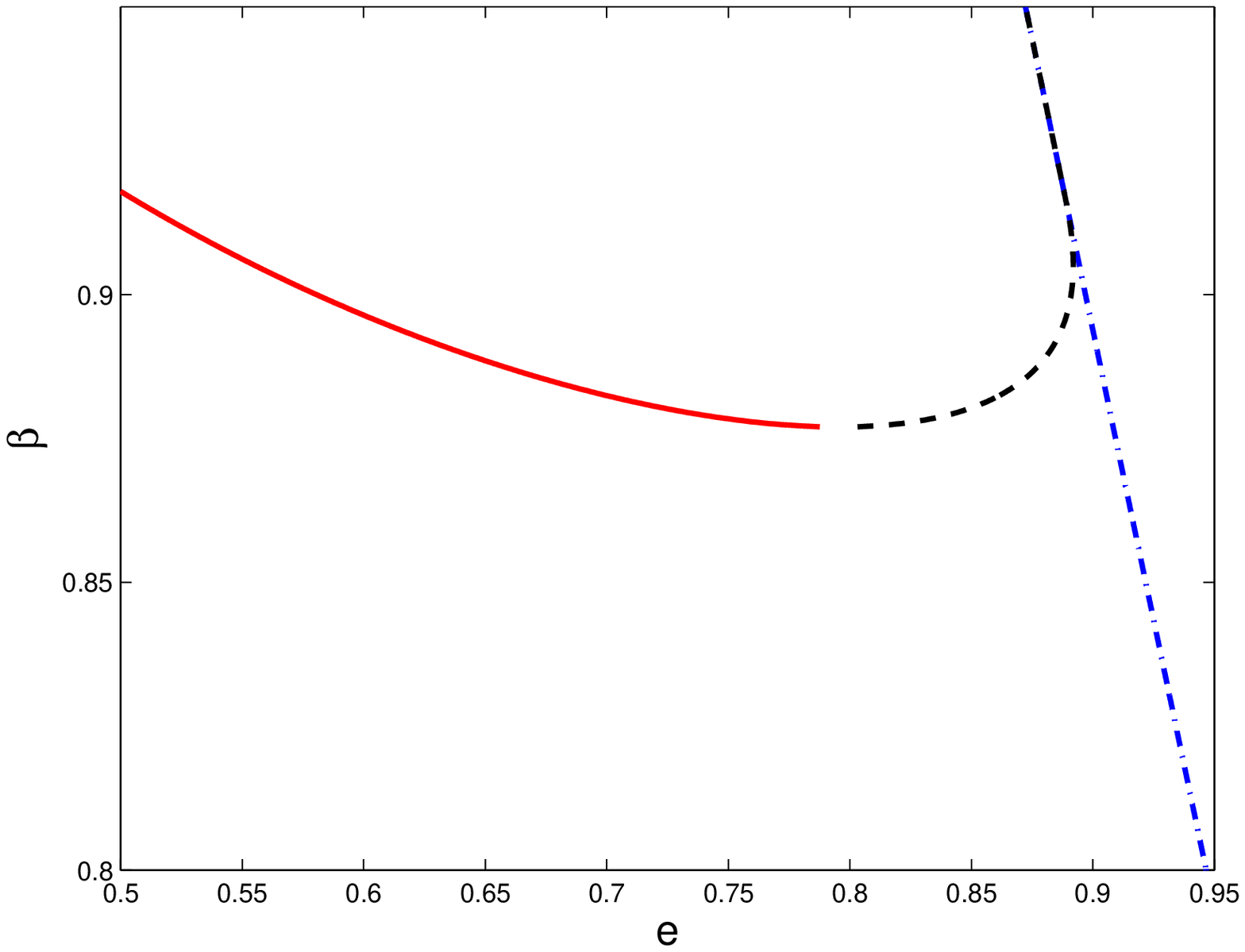}
\caption{\label{fig:Julien-s-y}Left: the three branches of the generalized 
free energy $\mathcal{F}$ as a function of the inverse temperature~$\beta$, 
for $k=4$. Right: the corresponding three branches for 
$\beta\left(e\right)$ in the microcanonical ensemble.}
\end{figure}

Plugging all the previous results in to Eq.~\ref{eq:F}, we obtain the
expression of the generalized free energy of the system for the general
$k$ case:
\begin{eqnarray}
\mathcal{F}\left(\beta\right)&=&-\ln\left[\left(p_{a}^{2}+p_{b}^{2}+p_{c}^{2}\right)+2\left(p_{a}p_{b}+p_{a}p_{c}+p_{b}p_{c}\right)e^{-\beta}\right]+\nonumber\\
&&+\frac{k}{2}\ln\left[\frac{1}{3}\left\{\left[p_{a}+\left(p_{b}+p_{c}\right)e^{-\beta}\right]^{k}+\right.\right.\\
&&+\left.\left.\left[p_{b}+\left(p_{a}+p_{c}\right)e^{-\beta}\right]^{k}+\left[p_{c}+\left(p_{a}+p_{b}\right)e^{-\beta}\right]^{k}\right\}\right]\nonumber
\end{eqnarray}

where the three densities $p_{a}$, $p_{b}$ and $p_{c}$ are solutions
of Eq.~\ref{eq:pabck}. Notice that this procedure does not necessarily
yield a unique ``free energy'' $\mathcal{F}\left(\beta\right)$;
rather, there is one value of $\mathcal{F}(\beta)$ for each solution
of the consistency equation~(\ref{eq:pabck}). We must then follow all
branches of the multi-valued function~$\mathcal{F}\left(\beta\right)$
to reconstruct the entropy~$S\left(e\right)$ through a generalized
inverse Legendre transform (see for instance~\cite{maragos} for a use
of this procedure in the context of signal processing):
\begin{equation}
S\left(e\right)=\beta e-\mathcal{F}\left(\beta\right)
\end{equation}
where:\[e\equiv\frac{\partial\mathcal{F}}{\partial \beta}\] can easily
be calculated numerically using finite differences. This is the final,
implicit, solution for the entropy $S\left(e\right)$.  In fig.
\ref{fig:Julien-s-y}, we plot the different solution branches of
$\mathcal{F}(\beta)$, and the inverse temperature
$\beta\left(e\right)$.  One clearly sees a negative specific heat
region, signaled by the presence of multiple function values for the
same energy.

\section{Comparison with numerical simulations}
\label{sec:numerical}

In this section we compare the analytical solution with the results
obtained through numerical simulations. Microcanonical simulations
were performed using Creutz~\cite{creutz} dynamics. During which, a
fictitious ``demon'' is introduced, carrying an energy $e_{demon}$. At
each step, a spin flip in the system is attempted, and the
corresponding energy change $\delta E$ is computed. If $\delta E<0$,
the move is accepted; if $\delta E>0$, the move is accepted only if
$e_{demon}\geq \delta E$.  In both cases $e_{demon}$ is then updated
so that the total energy $E+e_{demon}$ is kept constant; the energy of
the system~$E$ is then constant up to a $O(1/N)$.  For
long run times, the demon's energy reaches an exponential distribution
$P(e_{demon}=e)\propto \exp (-e/T_{\mu})$, from where one can compute
the corresponding microcanonical temperature $T_{\mu}=1/\beta_{mu}$ of
 our system:
\begin{equation}
\beta_{\mu}= \log\left[ 1+\frac{1}{\langle e_{demon}\rangle}\right] ~.
\end{equation}

\begin{figure}[htbp]
  \centering
    \includegraphics[width=10cm]{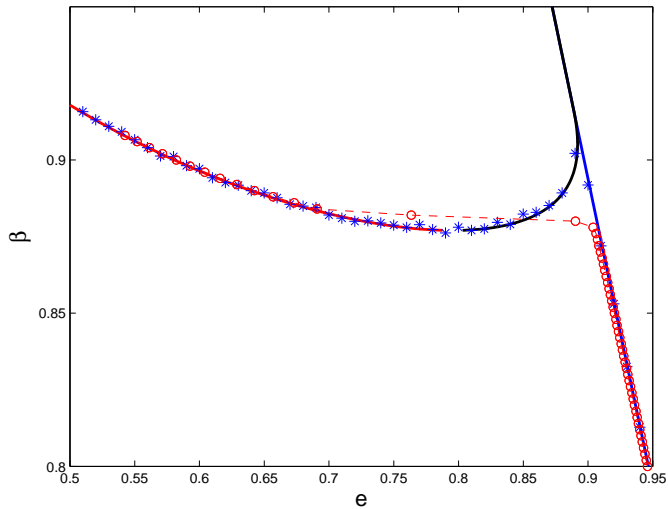}
    \caption{Comparison for the caloric curve $\beta\left(e\right)$
      between the analytical solution (solid lines), the Creutz
      dynamics results (stars), and the Metropolis simulations
      (circles) for $k=4$. The Creutz simulations were performed on
      networks with $N=40000$ sites, for $10^8$ ``Creutz steps'', and
      the results were averaged over $20$ network realizations. The
      Metropolis results were obtained using $50$ different networks
      with $N=10000$ nodes, by performing $10^{10}$ Monte-Carlo steps.
      In both cases, the size of the error bars is comparable to the
      size of the symbols.  }
\label{Fig:comparison}
\end{figure}

Results of the Creutz dynamics are plotted on
Fig.~\ref{Fig:comparison} and compared with the analytical solution of
the previous section. The agreement between the two is very good, with
the $\beta$~vs~energy curve clearly showing a region of negative
specific heat.

Finally, we performed canonical Metropolis\cite{metropolis} simulations and 
calculated the average energy in the temperature range where our results 
predict ensemble inequivalence. As expected, the canonical caloric curve obeys
 Maxwell's construction and clearly ``jumps over'' the region where the 
specific heat is negative.

\section{Conclusion and perspectives}
\label{sec:conclusion}

We have presented a complete canonical and microcanonical solution of
the 3-states Potts model on $k$-regular random graphs, and shown that
this toy model displays ensemble inequivalence.

There is little doubt that this result should generically apply to
models on different types of random graphs, such as Erd\"{o}s-R\'enyi
ones, among others. We also expect to observe ensemble inequivalence
on small world networks, since in these systems, the presence of
random long-range links should prevent the system from separating in
two different phases.

Beyond the inequivalence between microcanonical and canonical
statistical ensemble, non concave large deviation functions should be
expected for some properties on random graphs. Fig.~4 of~\cite{monasson}
gives an example of this. The present work provides an
example where the Large Deviation Cavity method allows to deal with
such a situation, and to compute the non concave part of the large deviation 
function. 

\begin{ack}
  We would like to acknowledge useful discussions with Stefan
  Boettcher, Matthew Hastings and Zolt\'an Toroczkai, and financial support 
from grant 0312510 from the Division of Materials Research at the National 
Science Foundation.
\end{ack}

\end{document}